\documentstyle[12pt]{article}
\newcommand{\mibitem}[1]{\bibitem{#1}}

\newcommand{\be}{\begin{equation}}
\newcommand{\ee}{\end{equation}}
\newcommand{\ba}{\begin{eqnarray}}
\newcommand{\ea}{\end{eqnarray}}
\newcommand{\bastar}{\begin{eqnarray*}}
\newcommand{\eastar}{\end{eqnarray*}}

\newcommand{\M}{{\cal M}}

\newcommand{\dop}{{\dot \phi}}
\newcommand{\X}{{\cal X}}

\newcommand{\bc}{{\bar c}}

\begin{document}
\begin{titlepage}
%%%%%%%%%%%%%%%%%%%%%%%%%%%%%%%%%%%%%%%%%%%%%%%%%%%%%%%
\begin{flushright}
UU-ITP 01-1995 \\
HU-TFT 95-04 \\
hep-th/9502055
\end{flushright}

\vskip 0.3truecm

\begin{center}
{ \large \bf
ON THE NUMBER OF PERIODIC CLASSICAL
\\ \vskip 0.2cm
TRAJECTORIES IN A HAMILTONIAN  SYSTEM
\\
}
\end{center}

\vskip 1.0cm

\begin{center}

{\bf Antti J. Niemi $^{*\dagger}$ } \\

\vskip 0.3cm

{\it Department of Theoretical Physics, Uppsala University \\

P.O. Box 803, S-75108, Uppsala, Sweden $^{\ddagger}$ \\

\vskip 0.4cm

and \\

\vskip 0.4cm

Research Institute for Theoretical Physics \\
P.O.Box 9,  FIN-00014 University of Helsinki, Finland \\
}

\end{center}

\vskip 3.0cm

\rm
\noindent
Periodic classical trajectories are of fundamental importance both in
classical and quantum physics. Here we develop path integral techniques to
investigate such trajectories in an arbitrary, not necessarily energy
conserving hamiltonian system. In particular, we present a simple
derivation of a lower bound for the number of periodic classical
trajectories.

\vfill

\vskip 0.4cm

\begin{flushleft}
\noindent
\rule{5.1 in}{.007 in}\\
$^{*}${\small E-mail: $~~$ {\small \bf
niemi@teorfys.uu.se}  \\
$^{\ddagger}$  permanent address\\
$^{\dagger}$ Supported by NFR Grant F-AA/FU 06821-308 and by
G{\"o}ran Gustafsson's Stiftelse} \\

\end{flushleft}

\end{titlepage}

\vfill\eject

\baselineskip 0.65cm

A periodic classical trajectory is a most important concept in Hamiltonian
mechanics, both at the classical and at the quantum level. A well known
example is given by the old Bohr quantization rule, which is a
condition on  periodic trajectories. Action-angle
variables which  parametrize periodic flow on the invariant torii  are
of fundamental importance in modern theory of classical and quantum
integrable models \cite{ludvig}. Furthermore, the only viable tool to
investigate  quantum chaos is the Gutzwiller (Selberg) trace formula
\cite{gutzwiller} which approximates energy spectrum of the quantum
Hamiltonian in terms of periodic classical trajectories.

For non-integrable models it is usually quite hard to obtain exact
results, and in practice one  often relies on  numerical investigations.
However, there are a number of important exact results.  Recently, the
Arnold conjecture on the number of periodic classical trajectories in an
arbitrary Hamiltonian system has been actively investigated
\cite{arnold1},  \cite{hofer}. The conjecture  states that on a compact phase
space an arbitrary but $T$-periodic, explicitly time dependent Hamiltonian
admits at least as many $T$-periodic classical trajectories as a function
has critical points. It  originates from problems in celestical mechanics
that date back to the turn of the century, and until now it
has only been verified in certain special cases \cite{hofer}.

A modern approach to the Arnold conjecture was developed by Floer
\cite{floer}, who formulated it in terms of infinite dimensional
Morse theory \cite{milnor}. His approach has been
extensively investigated \cite{hofer}, and it has also found applications in
other areas of theoretical physics, for example in topological field
theories and strings \cite{witten}.

In the present Letter we shall attempt a formulation of the
Arnold conjecture using quantum mechanical path integrals and the concept
of supersymmetry. Our approach is in principle applicable to an
arbitrary phase space without the compactness assumption \cite{floer},
\cite{hofer}, that  {\it a priori} excludes
several physically interesting examples.
We consider a general, $2N$ dimensional phase space $\M$ with local
coordinates $\phi^\mu$ and Poisson bracket $\{ \phi^\mu , \phi^\nu \}
= \omega^{\mu\nu}(\phi)$ \cite{arnold2}. We assume $\omega^{\mu\nu}$ is
nondegenerate so that we can introduce its inverse, the  symplectic two
form  $\omega_{\mu\nu} = - \omega_{\nu\mu} $. We are
interested in  Hamiltonian dynamics on $\M$, described by the
action
\be
S ~=~ \int_0^T dt \{ \ \vartheta_\mu {\dot \phi}^\mu - H (\phi ; t)  \ \}
\label{action}
\ee
where $\vartheta_\mu $ is the symplectic potential,
$\vartheta_\mu \dot \phi^\mu \sim p \dot q$ in local
Darboux coordinates \cite{arnold2}.  In general, the
Hamiltonian $H(\phi ;t)$ does not conserve energy {\it i.e.} it  depends
explicitly on time. We shall assume that the external forces are
$T$-periodic,  so that $H( \phi; t ) = H(\phi; t+T)$. The Hamilton's
equations of motion are
\be
{\dot \phi}^\mu - \omega^{\mu\nu} \partial_\nu H (\phi;t) ~ \equiv~ {\dot
\phi}^\mu -  \X^\mu_H (\phi;t) ~=~ 0
\label{equation}
\ee
and we are interested in estimating a lower bound for the number of
$T$-periodic classical trajectories, {\it i.e.} solutions to (\ref{equation})
with $\phi^\mu (0) = \phi^\mu (T)$.  It is sufficient to consider the
case where these solutions are non-degenerate.

If the Hamiltonian has no explicit time
dependence, the number of periodic
solutions can be bounded from below by the number of critical points of
$H(\phi)$: If we select the initial condition $\phi^\mu (0) = \phi^\mu_0$
where $\phi^\mu_0$ is a critical point of $H$, the configuration $\phi^\mu
(t) \equiv  \phi^\mu_0$ is trivially a $T$-periodic solution of
(\ref{equation}) for {\it any} $T$.  Consequently there are at least as many
$T$-periodic solutions as there are critical points in $H$, consistent with
the Arnold conjecture. On a compact $\M$ Morse theory
\cite{milnor} relates this lower bound to topology:
The number of critical points of $H$ is  bounded from below by the sum of the
Betti numbers $B_k = dim  H^k(\M , R)$, where $H^k(\M, R)$ is the $k$th
cohomology group.

However, if there are external time dependent forces the previous
argument fails since (time dependent) critical points of
$H(\phi;t)$ do not solve (\ref{equation}). The problem to estimate a
lower bound for the number of $T$-periodic solutions becomes now highly
nontrivial. According to Arnold \cite{arnold1}, \cite{hofer},  on a compact
$\M$ the previous estimate  should remain valid but this
conjecture remains unproven \cite{hofer}.

In the present Letter we  attempt a {\it quantum mechanical} path
integral approach to estimate a lower bound for the number of
$T$-periodic solutions for a general, explicitly time
dependent and $T$-periodic Hamiltonian. In particular, our approach should
also apply in the physically interesting case  of non-compact phase
spaces.  We  construct our estimate as follows:
\[
\#\{ \ T-periodic ~ solutions ~ to ~ (\ref{equation}) \ \} ~\equiv~
\sum\limits_{\delta S_{pbc}  = 0}
1 ~
\]
\[
=~ \int\limits_{\phi(T) = \phi(0)} [d\phi] \delta [ {\dot \phi}^\mu -
\X^\mu_H (t) ] \cdot \left| \ \det [ \delta^\mu_\nu \partial_t + \partial_\nu
\X^\mu_H (t) ] \ \right|  \]
\be
\geq ~ \left| \ \int [d\phi] \delta [ {\dot \phi}^\mu -
\X^\mu_H (t) ] \det [ \delta^\mu_\nu \partial_t + \partial_\nu \X^\mu_H
(t) ] \ \right|
\label{path1}
\ee
\be
= ~ \left| \ \sum\limits_{\delta S_{pbc}  = 0} {\rm sign} \{ \
\det [ \delta^\mu_\nu \partial_t + \partial_\nu \X^\mu_H (t) ] \ \} \
\right| \label{infmorse}
\ee
We identify the last sum as an infinite dimensional version of the sum that
appears in finite dimensional Morse theory \cite{milnor}, with $S$ the Morse
function. However, in general it is very difficult to extend results in
finite dimensional Morse theory to the present case: There is no minimum for
(\ref{action}), and periodic solutions of (\ref{equation}) are  saddle
points of  (\ref{action}) with an infinite Morse index. These problems have
been addressed by  Floer \cite{floer}, and we refer to \cite{hofer} for a
review.

Here we shall use the path integral in (\ref{path1}), to define the sum in
(\ref{infmorse}). We find that standard path integral techniques
provide a tool to evaluate (\ref{path1}), at least at the level of
heuristics  peculiar to  the Physics literature. In global Darboux
coordinates, such path integrals have been previously investigated  in
\cite{gozzi} and they are also intimately related to those in
stochastic quantization \cite{parisi}.

We shall employ two different versions
of (\ref{path1}). The first version is obtained, when we
introduce a commuting variable $p_\mu$ and two anticommuting variables
$c^\mu$, $\bc_\mu$ and write (\ref{path1}) as
\be
Z ~=~ \int[d\phi][dp][dc][d\bc]
\exp \{ i \int\limits_0^T p_\mu {\dot \phi}^\mu - \bc_\mu \dot c^\mu - p_\mu
\X^\mu_H  - \bc_\mu \partial_\nu \X^\mu_H  c^\nu \}
\label{path2}
\ee
The second version emerges when we introduce an
arbitrary non-degenerate  metric tensor $g_{\mu\nu}$ on the phase space
$\M$, and use a Gaussian representation of the $\delta$-function.
\[
Z ~=~ \lim_{\alpha \to 0}
\int[d\phi][dp][dc][d\bc] \exp \{ i \int\limits_0^T \frac{1}{2 \alpha^2}
g_{\mu\nu} ( \dop^\mu - \X^\mu_H) (\dop^\nu - \X^\mu_H) + g^{\mu\nu}
p_\mu p_\nu
\]
\be
+ \frac{1}{2 \alpha}
{\bar c}^\mu [ \ D_{\mu\nu} + D_{\nu\mu} + \Omega_{\mu\nu} + ( {\cal L}_H
g)_{\mu\nu} \ ] c^\nu   \ \}
\label{path3}
\ee
We have here added terms to the $c,  \bc$ part of the action which
are proportional to the argument of the
$\delta$-function in (\ref{path1}).  The operator
$$
D_{\mu\nu} =
g_{\mu\nu} \partial_t + g_{\nu\sigma} \dop^\rho \Gamma^\sigma_{\rho\mu}
$$
is
the pull-back of the covariant derivative on $\M$ with
$\Gamma^\sigma_{\mu\nu}$ the metric connection of $g_{\mu\nu}$, the matrix
$$
\Omega_{\mu\nu} = g_{\mu\rho} \partial_\nu \X^\rho_H -  g_{\nu\rho}
\partial_\mu \X^\rho_H
$$
is usually called the Riemannian momentum map of $H$
and ${\cal L}_H$ is the Lie derivative along the Hamiltonian vector field
$\X_H$. By construction, (\ref{path3}) is
independent of the metric $g_{\mu\nu}$ on $\M$. Furthermore, if the
Hamiltonian flow determines an isometry on $\M$ so that ${\cal L}_H g = 0$,
the operator that appears in the $c,  \bc$ part of the action becomes
antisymmetric.

We shall first use  (\ref{path2}) to show that our
lower bound  is {\it independent} of the
fact, that  time dependent external forces are present.  For this, we observe
that (\ref{path2}) has the canonical form of a Hamiltonian path
integral in Darboux coordinates. In particular, in (\ref{path2}) we identify
$p_\mu$ as a  canonical conjugate to $\phi^\mu$,  while
$c^\mu, \ \bc_\mu$ are mutually canonically conjugated variables.
Following \cite{oma1} we identify  $ c^\mu
\sim d\phi^\mu$, and $\bc_\mu$ as a basis of internal multiplication which is
dual to the $c^\mu$'s. We then define the  equivariant
exterior derivative
$$
d_H = d + i_H = c^\mu p_\mu + \X^\mu_H \bc_\mu
$$
and
find that the Hamiltonian in (\ref{path2}) is the corresponding Lie
derivative
$$
{\cal L}_H = d_H^2 = \{ d , i_H \}  =  p_\mu \X^\mu_H  +
\bc_\mu  \partial_\nu \X^\mu_H  c^\nu
$$
In particular, since $d = c^\mu
p_\mu$ is nilpotent we may view it as a BRST operator so that the path
integral (\ref{path2}) has the standard Fradkin-Vilkovisky form \cite{henne},
\be
Z_\Psi ~=~ \int[d\phi][d\lambda][dc][d\bc] \exp [ i \int\limits_0^T
p_\mu  {\dot \phi}^\mu - \bc_\mu {\dot c}^\mu - \{ d  ,  \psi \} \ ]
\label{FV}
\ee
According to the Fradkin-Vilkovisky theorem \cite{henne}, such path
integrals  are invariant under local variations of $\psi$: Since the
external forces are $T$-periodic we can Fourier expand
\be
H (\phi;t) ~=~ H_0(\phi) + \sum_{n\not= 0} H_n(\phi) e^{2\pi i \frac{t}{T} }
\label{hamexp}
\ee
where the coefficients $H_n(\phi)$ do not have any explicit
$t$-dependence. In the following we shall assume that the critical points
of $H_0$ which is the average of $H(\phi;t)$ over the period $T$, are
nondegenerate. In addition we shall assume that the $| \phi | \to \infty$
asymptotic behaviour of $H(\phi ;t)$ is determined by $H_0(\phi)$. On a
compact phase space this is of course not a restriction, but on a
non-compact phase space it means that the  global properties of the theory
are determined by $H_0$.   Later on we shall elaborate more on this
condition, at this moment it is sufficient to view it as a physically
reasonable assumption: For example, if $H_0$ corresponds to an anharmonic
oscillator with potential energy $V_0(q) =  \lambda_2 q^2 + \lambda_4 q^4$
this means that the  $H_n (\phi) \sim V_n(q)$ for $n\not= 0$
do not introduce terms that behave like ${\cal O}( q^4), ~  {\cal O} (q^5) ,
\ etc. $ for large $q$.

We set $\psi = i_H$ and $\psi_0 =
i_{H_0}$ so that $\psi_0$ depends only on the constant mode $H_0$, and
consider the change of variables
$$
\xi^\mu ~\to ~ \xi^\mu +  \{ d ,
\xi^\mu \} \cdot \epsilon \int_0^T ( \psi - \psi_0 )
$$
in (\ref{FV})
where $\xi^\mu$ denotes the variables $\phi^\mu$, $p_\mu$, $c^\mu$ and
$\bc_\mu$. We find \cite{henne} that its {\it only} effect is the shift $
\psi ~\to~  (1-\epsilon) \psi + \epsilon \psi_0 $ in (\ref{FV}).
Consequently  (\ref{FV}) does {\it not} depend on the modes
$H_n$ for $n\not= 0$, and in particular the lower bound  (\ref{path1})
coincides with the lower bound obtained with the  $t$-independent
$H_0(\phi)$.

We now consider the path integral (\ref{path3}): According to
the previous arguments it is invariant under local
variations of $H(\phi;t)$. Hence it is sufficient to evaluate (\ref{path3})
for the dominant, $t$-independent mode $H_0(\phi)$ only. For this, we
introduce the following  Fourier-expansions
$ \phi^\mu(t) = \phi^\mu_0 + \delta \phi^\mu (t) \ , \  ... \ , \ \bc_\mu (t)
= \bc_{\mu 0} + \delta \bc_\mu (t)$
where $ \phi^\mu_0 \ , \ ... \ , \ \bc_{\mu 0}$ are the $t$-independent
constant modes, and $ \delta  \phi^\mu (t) \ , \ ... \ , \  \delta \bc_{\mu}
(t)$ are the $t$-dependent fluctuation modes. We  change
variables $\delta \phi^\mu  \to \alpha \delta \phi^\mu ~$, $ \delta c^\mu
\to \sqrt{\alpha} \delta c^\mu  ~$ and $ \delta \bc_\mu  \to \sqrt{\alpha}
\delta \bc_\mu$.  The corresponding super-Jacobian in (\ref{path3}) is
trivial and in the $\alpha \to 0$ limit (\ref{path3}) localizes to the
critical points of $H_0(\phi)$ so that  (\ref{path1}) yields us
the following lower bound
\be
\# \{ \ T \!-\!periodic ~ solutions ~ to ~ (\ref{equation}) \ \} ~\geq~
\left|  \ \sum\limits_{ dH_0 = 0 }  { \rm sign \ det \ } \|
\partial_{\mu\nu} H_0 \| \ \right|
\label{estimate}
\ee
for the number of $T$-periodic  solutions to (\ref{equation}).  Notice
that  we have not assumed compactness of the phase space, and we have only
used standard path integral techniques that are expected to be valid  quite
generally. At this point  we can also better understand our condition,
that for all $t$ the large-$| \phi |$ asymptotic behaviour of $H(\phi ; t)$
is determined by $H_0$:  Morse theory \cite{milnor} connects the invariance
of the lower bound in (\ref{estimate}) to the asymptotic  large-$| \phi |$
behaviour of $H_0$. If this is changed by some $H_n$ ($n\not=
0$) in (\ref{hamexp}), the argument by Fradkin and Vilkovisky fails
to  remove  $H_n$ since the corresponding variation $\psi \to \psi + \delta
\psi_n$  is  "large". For a {\it compact} phase space this is not an issue.
Indeed, we can  relate (\ref{estimate}) to the manifestly
$H$-independent Euler characteristic of $\M$ as follows: If we define the
nilpotent BRST operator
\[
Q ~=~ c^\mu
{\partial \over \partial \phi^\mu} + p_\mu  {\partial \over \partial
\bc_\mu}
\]
since ${\cal L}_H = \{ Q , \X^\mu_{H_0} \bc_\mu \}$ we can use it to
represent (\ref{path2}) in the corresponding Fradkin-Vilkovisky form
(\ref{FV}). In particular, the path integral is invariant under local
variations of the form  $\X^\mu_{H_0} \bc_\mu \equiv \psi ~\to~ \psi +
\delta \psi$.  We introduce the  conjugation $Q ~\to~ e^{-\theta} Q
e^\theta$ and select $\theta = - \Gamma^\rho_{\mu\nu} c^\mu \bc_\rho
\pi^\nu$ where $\pi^\nu p_\mu = \delta^\nu_\mu$ and $\Gamma^\rho_{\mu\nu}$
is the connection for some metric $g_{\mu\nu}$ on $\M$. We find \be
Q ~=~ c^\mu \partial_\mu + (p_\mu + \Gamma^\rho_{\mu\nu} c^\nu \bc_\rho)
{\partial \over \partial \bc_\mu } + (\Gamma^\rho_{\mu\nu} p_\rho c^\nu
- \frac{1}{2} {R^\sigma}_{\mu\rho\nu} c^\nu c^\rho \bc_\sigma) \pi^\mu
\label{sigmaQ}
\ee
where ${R^\sigma}_{\mu\rho\nu}$ is the Riemann curvature tensor on $\M$,
and we recognize (\ref{sigmaQ}) as the supersymmetry generator of the
standard quantum mechanical N=1 supersymmetric nonlinear $\sigma$-model.
Our conjugation leaves the path integral measure invariant, but shifts
$ p_\mu ~\to~ p_\mu  + c^\nu \Gamma^\rho_{\mu\nu} \bc_\rho$. Selecting
$\psi = g^{\mu\nu} p_\mu \bc_\nu$
we find  that the corresponding (\ref{path2}) localizes to
the Euler characteristic of the phase space $\M$,
\[
Z ~=~ \int_\M {\rm Pf} \ [
\frac{1}{2}{R^\mu}_{\nu\rho\sigma} c^\rho c^\sigma ]
\]
For {\it compact} phase spaces we then conclude that
(\ref{estimate}) is equivalent to
\be
\#\{ \ T \!-\!periodic ~ solutions ~ to ~ (\ref{equation}) \ \} ~\geq~ \left|
\ \int_\M {\rm Pf} \ [  \frac{1}{2}{R^\mu}_{\nu\rho\sigma} c^\rho c^\sigma ]
\ \right| ~=~ \left| \ \sum_k  (-)^k H^k(\M , R) \ \right|
\label{euler1}
\ee
Notice that if all $B_{2k+1} = 0$ and there are no
degeneracies,  (\ref{euler1}) is exactly the lower bound conjectured by
Arnold \cite{arnold1}, \cite{hofer}. Notice also that on a {\it non}-compact
$\M$ we can not use the Fradkin-Vilkovisky argument to set $\X^\mu_{H_0}
\bc_\mu \to g^{\mu\nu} p_\mu \bc_\nu$ in the path integral, since we can not
change the asymptotic behaviour of $H_0$ at infinity. Indeed, for a
non-compact phase space such as $R^{2n}$ the {\it r.h.s.} of (\ref{euler1})
vanishes, while (\ref{estimate}) is in general  nontrivial.

If the Hamiltonian flow generated by $H$ defines an {\it isometry} of the
metric tensor $g_{\mu\nu}$ so that ${\cal L}_H g = 0$, the operator that
appears in the $c,\bc$ part of the action in (\ref{path3}) is
antisymmetric. Using standard properties of antisymmetric matrices
we then conclude that the signs of our determinants coincide, and
if we select orientation properly we have ${ \rm sign \  det \
} \| \partial_{\mu\nu} H_0 \| = +1$ for all terms in the {\it r.h.s.} of
(\ref{estimate}). In this way we obtain that for {\it isometric}
Hamiltonians
\be
\#\{ \  T
\!-\!periodic ~ solutions ~ to ~ (\ref{equation}) \ \} ~=~  \sum_{
dH_0 = 0}  1
\label{euler2}
\ee
if $T$ is selected so that there are no degeneracies.
Since for a {\it compact} phase space the {\it r.h.s.}
coincides with the Euler character, we have verified that on
compact $\M$ isometric Hamiltonians exist only if odd
Betti numbers $B_{2k+1}$ vanish. In standard Morse theory such isometric
$H$ are called perfect Morse functions, and we have established that this
concept  generalizes to the infinite dimensional context.

For a compact phase space we can view (\ref{euler1}), (\ref{euler2})
as a path integral proof of the  Lefschetz fixed
point theorem \cite{eguchi} for an
arbitrary and for an isometric Hamiltonian, respectively:
According to Lefschetz, if $F: \ \M \to \M$ is a smooth map on a compact
Riemannian manifold $\M$ which is homotopic to the identity and admits only
isolated fixed points $F[x] = x$,  the number of these fixed points is
bounded from below by the Euler character of $\M$. Furthermore, if $F$ is an
isometry this number {\it coincides} with the Euler character. In the present
context, Hamilton's equations of motion  define a smooth mapping  $\ F: \M
\to \M$  by $F[ \phi] \equiv F[\phi(t=0) ] ~\to~ \phi (T)$. The time
evolution ensures that this mapping is homotopic to the identity, while the
fixed point condition $F[\phi] = \phi$ selects the $T$-periodic classical
trajectories $\phi(0) = \phi(T)$.
\vskip 0.3cm

Finally, we observe that the present construction can be given a
very suggestive supersymmetry interpretation: We have already
pointed out the similarity between  (\ref{path1}) and the
path integrals that appear in the Parisi-Sourlas approach to stochastic
quantization \cite{parisi}. Indeed, if we introduce the
supercoordinates $\theta$ and $\bar\theta$ and define the
superfield $\Phi^\mu(\theta,\bar\theta) = \phi^\mu +
c^\mu \theta + \omega^{\mu\nu}\bc^\nu \bar\theta + \omega^{\mu\nu} p^\nu
\theta\bar\theta $, as in \cite{parisi} we
find that in this superspace the action in (\ref{path2}) admits
a functional form which is {\it identical}
to its functional form (\ref{action}) in the original phase space,
\[
\int\limits_0^T dt \{ \ p_\mu\dop^\mu - \bc_\mu \dot c^\mu - p_\mu \X^\mu_H -
c^\mu \partial_\mu \X^\nu_H \bc_\nu \ \} ~=~ \int\limits_0^T dt d\theta
d\bar\theta \ \{ \ \vartheta_\mu [\Phi] {\dot \Phi}^\mu - H [\Phi] \ \}
\]
Indeed, we argue that conceptually our localization formulas are nothing but
generalizations of the Parisi-Sourlas integral \cite{parisi},
$$
\int d^2x d\theta d\bar\theta \ F( x^2 + \bar\theta \theta) ~=~ F(0)
$$

\vskip 0.5cm
We shall now consider two {\it examples} \cite{losev}: Our first example
is the canonical realization of classical spin. It employs phase space
coordinates that are not of the Darboux form. The phase space is the
two-sphere $S^2$ which is a compact manifold, and the classical action is  $
S ~=~ j \int_0^T \cos\theta  \dot \phi - cos \theta \ $.
The equations of motion (\ref{equation}) becomes $\sin \theta \cdot
\dot\theta = 0$ and $\sin \theta (\dot\phi -1 ) = 0$. For $T=2\pi n$ we have
periodic solutions for {\it any} initial condition on $S^2$. For other
values of $T$, the only periodic solutions are $\theta = 0$ and $\theta =
\pi$. These are the two critical points of the Hamiltonian  $H_0 = j
cos\theta$, both with a Morse index ${\rm sign} \det || \partial_{\mu\nu} H
|| = +1$.

Suppose now $T \not= 2\pi n$ so that the only $T$-periodic classical
trajectories are the two critical ponts at $\theta = 0 $ and $\theta = \pi$.
We add an arbitrary time dependent force with period $T$ described by some
Hamiltonian $H_1(\phi,\theta ; t)$, for example a time dependent magnetic
field. Now the critical points at $\theta = 0$ and $\theta = \pi$ cease to
solve the classical equations of motion but according to (\ref{estimate})
there are {\it at least} two $T$-periodic solutions. This result also
follows directly from the Lefschetz fixed point formula \cite{eguchi}. It is
also consistent with the Arnold conjecture, since the Betti numbers for
$S^2$ are $B_0 = B_2 = 1$. In particular, the Hamiltonian flow of $H_0  = j
\cos \theta \sim j \hat z $ defines an isometry of  $S^2$ with its standard
metric. If the time-dependence only defines a rotation of $\hat z$ in $R^3$
so that for all $t$ we have an isometry, then (\ref{euler2}) implies that
there are {\it exactly} two $T$-periodic classical trajectories for {\it
any} value $T \not= 2\pi n$.

The second example  corresponds to a non-compact phase space. Hence the
Lefschetz fixed point formula is not applicable, but our result
(\ref{estimate}) should remain valid: We consider a charged particle of mass
$m$ constrained to move on the two-sphere $S_r^2$ with radius $r$ in three
dimensional Euclidean space, with equator on the $x-y$ plane and in a
gravitational potential $V(x,y,z) = mg z$. The phase space is now the
cotangent bundle $R^2 \times S_r^2$ which is noncompact. In the absence of
other interactions and for a generic $T$, the only $T$-periodic classical
trajectories are those corresponding to the particle sitting at rest at $z =
\pm r$. These are both critical points of $H$ with Morse index $+1$. If we
now add  arbitrary but $T$-periodic electric and magnetic fields, and if we
also deform (but dont break) the sphere and toss it around in
three-space in an arbitrary but $T$-periodic manner, according to our
estimate (\ref{estimate}) there are still {\it at least} two $T$-periodic
classical trajectories.

\vskip 0.8cm
In conclusion, we have investigated the space of $T$-periodic classical
trajectories in a general, explicitly time dependent and
$T$-periodic Hamiltonian system defined in an arbitrary phase space. In
particular, we have found that at least formally, lower bounds of the
Lefschetz type can be generalized to estimate the number of periodic
classical trajectories also for non-compact phase spaces. Numerical
investigations of some simple examples could provide interesting insight to
the behaviour of such solutions.

\vskip 0.3cm
\noindent
We thank M. Blau, V. Fock, A. Losev and O. Viro for discussions.

\end{document}